# Conditions for exact equivalence of Kaluza-Klein and Yang-Mills theories

Frank Reifler and Randall Morris

Lockheed Martin Corporation MS2 137–205
199 Borton Landing Road
Moorestown, NJ  08057, U.S.A.

**Abstract:**  Although it is well known that Kaluza-Klein and Yang-Mills theories define equivalent structures on principal bundles, the general conditions for equivalence of their Lagrangians have not been explicitly stated.  In this paper we address the conditions for equivalence.  The formulation of these conditions is based on previous work in which the Dirac and Einstein equations were unified in a tetrad formulation of the Kaluza-Klein model.  This Kaluza-Klein model is derived from mapping a bispinor field $\Psi$ to a set of $SL(2,R) \times U(1)$ gauge potentials $F_\alpha^K$ and a complex scalar field $\rho$.  (A straightforward derivation of this map using Hestenes' tetrad for the spin connection in a Riemannian space-time is included in this paper.) Investigation of this Kaluza-Klein model reveals two general conditions for establishing an exact equivalence between Kaluza-Klein and Yang-Mills theories.  The first condition is that only horizontal vector fields occur in the Kaluza-Klein Lagrangian.  The second is that the scalar curvature be restricted to a sum over horizontal sectional curvatures.  We conclude that all known fields (including fermion fields) can then be represented as components of a Kaluza-Klein metric together with scalar fields.

## 1.  Introduction

It is well known that Kaluza-Klein and Yang-Mills theories define equivalent structures on principal bundles.  However, the general conditions for equivalence of their Lagrangians have not been explicitly stated [1] – [18].  In this paper we address Kaluza-Klein manifolds of the form M = X × G, where X is a four-dimensional space-time, and G is an arbitrary Lie group (gauge group) with a fixed right-invariant metric that, to conform to Yang-Mills theory, does not depend on the space-time.  For the corresponding Yang-Mills theory, M is then a trivial principal G-bundle.  Defining a Yang-Mills connection on the principal bundle M then defines the Kaluza-Klein metric uniquely on M given the following three requirements [2]:

(1) The horizontal and vertical subspaces in the tangent bundle of M, defined by the Yang-Mills connection, are orthogonal with respect to the Kaluza-Klein metric.
(2) The space-time metric is the Kaluza-Klein metric restricted to the horizontal subspaces.
(3) The Kaluza-Klein metric restricted to the vertical subspaces is the right-invariant metric chosen on the gauge group G.

Conversely, given a Kaluza-Klein metric on M, we can derive a Yang-Mills connection as follows. The Kaluza-Klein metric decomposes the tangent bundle of M into two orthogonal subbundles, that is, into the vertical subbundle of all tangent vectors of G and the horizontal subbundle of all vectors orthogonal to the tangent vectors of G. The horizontal subbundle inherits the right invariance of the Kaluza-Klein metric, and hence defines a unique Yang-Mills connection on the principal bundle M. Thus, there is a one-to-one correspondence between the Kaluza-Klein metrics satisfying conditions (1), (2), and (3), and Yang-Mills connections on the manifold M = X × G. In this sense Kaluza-Klein and Yang-Mills theories define equivalent structures.

However, the Lagrangian in current Kaluza-Klein theory differs from the Yang-Mills Lagrangian. Generally, a Kaluza-Klein Lagrangian has terms not present in the Yang-Mills Lagrangian. For example, vertical as well as horizontal vector fields act on the scalar fields in a Kaluza-Klein Lagrangian, whereas only horizontal vector fields act on the scalar fields in a Yang-Mills Lagrangian [6], [14]. Also, the scalar curvature (i.e. the sum over all sectional curvatures of M) in a Kaluza-Klein Lagrangian, produces a cosmological constant that is not present in a Yang-Mills Lagrangian [4], [5], [7]. Furthermore, if the right-invariant metric on the gauge group G is not bi-invariant, unwanted terms occur in the Kaluza-Klein Lagrangian [2], [10]. Thus for the Kaluza-Klein and Yang-Mills Lagrangians to be equal, additional conditions are required.

In this paper we will derive two conditions for establishing equality of Kaluza-Klein and Yang-Mills Lagrangians based on recent work in unifying the Dirac and Einstein equations in a tetrad formulation of Kaluza-Klein theory [19] – [22]. The first condition is that only horizontal vector fields occur in the Lagrangian. The second condition is that the scalar curvature in the Kaluza-Klein Lagrangian be restricted to the horizontal subbundle. That is, the sum over sectional curvatures of M in the Kaluza-Klein Lagrangian must be restricted to horizontal sectional curvatures.

Note that in Yang-Mills theory, scalar fields transform covariantly under the gauge group G. No additional assumption is required for this in Kaluza-Klein theory, since covariant scalar fields are special solutions of the Kaluza-Klein equations (see Section 3).

Given the condition that we restrict to horizontal vector fields and horizontal curvature in the Kaluza-Klein Lagrangian, we will show in Section 3 that the Kaluza-Klein and Yang-Mills Lagrangians are equivalent, using as an example the unification of gravity and fermions [19] – [22]. It is common in Kaluza-Klein theory to consider higher dimensional bispinor fields defined on the Kaluza-Klein manifold that are not unified with the Kaluza-Klein metric, with supernumerary boson degrees of freedom in addition to the fermion degrees of freedom [6], [13], [17], [18]. This paper uses a different approach, whereby four-dimensional gravitational and fermion fields are faithfully represented by components of a Kaluza-Klein metric together with a scalar field. In this formulation, the Kaluza-Klein Lagrangian precisely reduces to the usual Einstein-Dirac Lagrangian. Also, in this formulation, there are no superfluous fields or supernumerary degrees of freedom (see

Section 3). This parsimonious feature has not been achieved in previous theories of fermions in a curved space-time.

That is, in order to define bispinors, reference tetrad fields, or their equivalent, must be defined on the space-time manifold [23] – [31]. However, only ten of the sixteen components of a tetrad field describe gravity. The remaining six components are supernumerary boson fields in current gravitational theories. In the Kaluza-Klein tetrad model, the tetrads, which simply define the Kaluza-Klein metric, describe both fermions and gravity without superfluous degrees of freedom [19] – [22]. For the rest of this introduction we will further describe the unification of the Einstein-Dirac equations to support the argument that all known fields (including fermion fields) can be unified as components of a Kaluza-Klein metric together with covariant scalar fields.

Using geometric algebra, Hestenes showed in 1967 that a bispinor field $\Psi$ on a Minkowski space-time is equivalent to an orthonormal tetrad of vector fields $e_a^\alpha$ together with a complex scalar field $s$, and that fermion plane waves can be represented as rotational modes of the tetrad [32]. As stated by Takahashi, who in 1983 derived the tensor form of the Dirac bispinor Lagrangian in terms of Hestenes' tetrad and scalar field, and who is well known for his work in quantum field theory [33] – [35], "…. a tetrad in a Minkowski space implies the existence of a spinor, and the [spinor] orthogonality and completeness conditions are automatically satisfied, when the tetrad is expressed in terms of the spinor." [33] These orthogonality and completeness conditions give rise to the oscillator modes that lead directly to fermion creation and annihilation operators [36]. Indeed, the oscillator modes of a bispinor field are precisely the oscillator modes of a tetrad field. From this point of view, as recognized by both Hestenes and Takahashi, there can be no difference between the bispinor and tensor theories for establishing the quantum field theory of fermions.

Hestenes explains the relation between the bispinor field $\Psi$ and the tensor fields $e_a^\alpha$ and $s$ as follows: "One sometimes finds in the literature the cryptic assertion that spinors are more fundamental than vectors, because a spinor is a kind of square root of a vector. This view reveals an incomplete grasp of the geometric meaning of spinors. Spinors cannot be defined without reference to vectors. …. The geometric meaning of a spinor is operational: it transforms one vector (or frame of vectors) into another." (See [37], page 1023.) In this quotation of Hestenes, the "initial" frame of vectors refers to an arbitrary reference frame, such as a laboratory frame, defined on a Minkowski space-time, and the transformed frame of vectors refers to Hestenes' tetrad field $e_a^\alpha$, determined by the bispinor field $\Psi$ itself. In saying that spinors are not "more fundamental than vectors", Hestenes argues that all physical properties of bispinors are manifested in the tetrad field $e_a^\alpha$ and scalar field $s$.

However, there are two assumptions which underlie Hestenes' and Takahashi's claim that Hestenes' tensor fields are equivalent to bispinors and that they describe fermions. The first assumption is that global tetrad fields exist on the space-time. This assumption is satisfied if the space-time is four-dimensional, admits spinor structure, and is non-compact [38] – [44]. On such space-times, spinor structures and homotopy classes of global tetrad fields are synonymous [40]. The second assumption, required for a consistent interpretation within quantum mechanics, is that we restrict to solutions of the tensor Dirac equation having "unique continuation". That is, for any observer, the history of a tetrad field in the past must uniquely determine its evolution into the future [45] – [51]. Non-unique continuation of solutions occurs in other areas of physics, such as in fluid dynamics, but is not considered to be appropriate for quantum mechanics. Note that there can be a bispinor field having unique continuation, whose tetrad field cannot be uniquely predicted into the future [45] – [47]. However, in a Minkowski space-time, continuation of the tetrad field is unique if we restrict to physically realizable bispinor solutions of the Dirac equation whose energy spectrum is bounded from below [48] – [51]. For example, the thought experiment [45] proposed by Y. Aharonov and L. Susskind to observe the sign change of bispinors under $2\pi$ rotations is not physically realizable, because the bispinor field would have to have an unbounded energy spectrum stretching from negative infinity to positive infinity. This could only be achieved by superposition of particle and anti-particle solutions, which is forbidden by a superselection rule in quantum mechanics [36].

In this paper, we treat a bispinor field $\Psi$ as a classical field, rather than as a quantum mechanical wave function as in the discussion above. It has been shown that every bundle of spin frames on a non-compact four-dimensional space-time with spinor structure is trivial [40], [44]. Hence, on any open subset U of the space-time where a reference tetrad is defined, bispinor fields $\Psi$ can be simply expressed as maps $\Psi : U \to C^4$, where $C^4$ is a four-dimensional complex vector space [40], [44]. Furthermore, in a non-compact four-dimensional space-time with spinor structure, the Einstein-Dirac equations depend only on a tetrad and a scalar field [52], [53]. This can be demonstrated by an appropriate choice of reference tetrad. The appropriate choice is provided by Hestenes' orthonormal tetrad of vector fields, denoted as $e_a^\alpha$, where $\alpha = 0, 1, 2, 3$ is a space-time index and $a = 0, 1, 2, 3$ is a tetrad index [32]. Relative to this special reference tetrad, a bispinor field $\Psi$ is "at rest" at each space-time point and has components given as follows: (See Section 2.)

$$\Psi = \begin{bmatrix} 0 \\ \text{Re}[\sqrt{s}] \\ 0 \\ -i\,\text{Im}[\sqrt{s}] \end{bmatrix} \tag{1.1}$$

where s is a complex scalar field defined in Section 2 by formula (2.4). Note that Hestenes' tetrad $e_a^\alpha$ and the complex scalar field $\sqrt{s}$ are smoothly defined locally in open regions about each space-time point where s is nonvanishing. In each open region, the Dirac bispinor Lagrangian depends only on the reference tetrad $e_a^\alpha$ and on the bispinor field $\Psi$ [23] – [30]. We show in Section 2, using formula (1.1), that the Einstein-Dirac Lagrangian can be expressed entirely in terms of the tensor fields, $e_a^\alpha$ and s, once Hestenes' tetrad has been chosen as the reference.

Whenever $\Psi$ vanishes, both s and its first partial derivatives vanish. Setting s and its first partial derivatives to zero in the tensor form of Dirac's bispinor equation shows that $e_a^\alpha$ can be chosen arbitrarily at all space-time points where $\Psi$ vanishes. Thus, all aspects of Dirac's bispinor equation are faithfully reflected in the tensor equations (see Section 2). Since the tetrad $e_a^\alpha$ is unconstrained by the Dirac equation when $\Psi$ vanishes, a gravitational field exists even if the fermion field vanishes. In Section 3 we show that the gravitational field $g_{\alpha\beta}$ and the bispinor field $\Psi$ (which together have 10 + 8 = 18 real components), are represented accurately by Hestenes' tensor fields $e_a^\alpha$ and s (which also have 16 + 2 = 18 real components) [19] – [22].

Hestenes' tetrad has been of interest for other applications. Zhelnorovich used Hestenes' tetrad together with the bispinor field at rest as in formula (1.1) to derive spatially flat Bianchi type I solutions of the Einstein-Dirac equations [52], [53]. Hestenes' tetrad in this application has the advantages of reducing the number of unknowns by six and of not requiring special symmetry directions for choosing the tetrad, which considerably simplifies the Einstein-Dirac equations for non-diagonal metrics and makes it possible to obtain new exact solutions [52], [53].

It might seem that Hestenes' tensor fields do not lead to a well posed initial value problem when isolated parts of a bispinor field, with disjoint (closed) supports in a Minkowski space-time, are rotated 360 degrees relative to one another [45] – [47]. However, such isolation is not possible because physical bispinor fields with energy bounded from below have closed supports filling all of space-time [48] – [51]. For example, in realistic beam splitting experiments [36], [46], [47], the beams always overlap, since bispinor fields with energy bounded from below cannot vanish identically on any open subset of the space-time [48] – [51]. For the special case of freely propagating bispinor fields, Thaller and Thaller state that " free Dirac wave packets with positive energy cannot be localized in any finite region and even cannot vanish on any open region of space" [49]. Hence, the tensor fields $e_a^\alpha$ and s determine a physical bispinor field $\Psi$ uniquely, up to a single unobservable sign [50], [51]. Thus, the two possible signs for $\Psi$, manifested by the $\sqrt{s}$ in formula (1.1), have no physical consequence when representing $\Psi$ with the tensor fields $e_a^\alpha$ and s. [36]

Note that from a quantum mechanics standpoint, the relative 360 degrees rotation of fermion particles in disjoint regions of space-time is not directly observed in current experiments. As stated by Byrne for double beam neutron experiments: "The relative rotation of the spins in the two beams and the interference pattern are mutually incompatible observables" that cannot be jointly measured [46]. "Although one can measure the rotation of neutrons along one path relative to those along the other, to do so would destroy the interference pattern." [47] In this paper, we treat a bispinor field $\Psi$ as a classical field, rather than as a quantum mechanical wave function describing a point particle, as in the arguments presented by Byrne and Silverman [46], [47]. Note, however, that Hestenes' tensor fields $e_a^\alpha$ and s faithfully represent all observable properties of a bispinor field $\Psi$ in either interpretation.

The Kaluza-Klein tetrad model is based on a constrained Yang-Mills formulation of the Dirac theory [19], [20]. In this formulation Hestenes' tensor fields $e_a^\alpha$ and s are mapped bijectively onto a set of $SL(2,R) \times U(1)$ gauge potentials $F_\alpha^K$ and a complex scalar field $\rho$. Thus we have the composite map $\Psi \to (e_a^\alpha, s) \to (F_\alpha^K, \rho)$ (see Section 2). The fact that $e_a^\alpha$ is an orthonormal tetrad of vector fields imposes an orthogonal constraint on the gauge potentials $F_\alpha^K$ given by

$$F_\alpha^K F_{K\beta} = |\rho|^2 g_{\alpha\beta} \qquad (1.2)$$

where $g_{\alpha\beta}$ denotes the space-time metric. The gauge index K = 0, 1, 2, 3 is lowered and raised using a gauge metric $g_{JK}$ and its inverse $g^{JK}$ (see Section 2). Repeated indices are summed. In Section 2 we show, for massive fermions, that putative differences between the bispinor field $\Psi$ and the tensor fields $F_\alpha^K$ and $\rho$ cannot be observed in experiments. We further show that via the map $\Psi \to (F_\alpha^K, \rho)$ the Dirac bispinor Lagrangian (2.3) equals a constrained Yang-Mills Lagrangian (2.18) in the limit of an infinitely large coupling constant, denoted as $g_0$ (see Section 2).

In the Kaluza-Klein formulation of the tensor Dirac theory, we map the fermion field $(F_\alpha^K, \rho)$ to a tetrad of horizontal vector fields $v_K$ and a complex scalar field, also denoted as $\rho$, on a smooth manifold $M = X \times G$, where X is a space-time and $G = SL(2, R) \times U(1)$ (see Section 3). The horizontal tetrad $v_K$ together with a (fixed) vertical basis of right-invariant vector fields on G determines a Kaluza-Klein metric denoted as $\langle , \rangle$, a volume form denoted as $d\gamma$, and also a curvature two-form denoted as $R( , )$, on M (see Section 3). The unified action S for the gravitational and fermion fields is given by

$$S = \int L \, d\gamma \qquad (1.3)$$

where the unified Lagrangian, L, is (see Section 3):

$$L = \frac{1}{16\pi\kappa_0} R_v + \frac{1}{g_0} \overline{v_K(\rho+\mu)} \, v^K(\rho+\mu) \qquad (1.4)$$

where $\kappa_0$ is Newton's gravitational constant, $g_0$ is the Yang-Mills coupling constant referred to previously, and $\mu$ is a complex scalar Higgs field defined on M which generates the fermion mass. In formula (1.4), we employ the sum of sectional curvatures restricted to the horizontal subspaces spanned by the tetrad $v_K$:

$$R_v = \sum_{J=0}^{3} \sum_{K=0}^{3} \langle R(v_J, v_K) v^J, v^K \rangle \qquad (1.5)$$

Also in formula (1.4), only the horizontal vector fields $v_K$ act on the scalar fields $\rho$ and $\mu$. By formulating the Kaluza-Klein Lagrangian (1.4) with the horizontal tetrad $v_K$, the orthogonal constraint (1.2) is eliminated (see Section 3).

The limit on the Yang-Mills coupling constant $g_0$ also has a geometric significance in the Kaluza-Klein tetrad model, in that as $g_0$ becomes infinitely large, as required to obtain the usual Einstein-Dirac equations from the Lagrangian (1.4), the radius of the higher compact dimensions in the Kaluza-Klein model becomes vanishingly small [21]. This can be seen from the following argument. In the Lagrangian (1.4) the constants $g_0$ and $\kappa_0$ are functions of two "fundamental" constants $\delta_0$ and $\lambda_P$, where $\delta_0$ is a radius which characterizes the size of the higher compact dimensions of the Kaluza-Klein manifold M, and $\lambda_P$ is the Planck length. In Section 3 we show that

$$\delta_0 = \left(\frac{8\pi}{g_0^3}\right)^{1/2} \lambda_P \qquad (1.6)$$

Thus, in the limit required to obtain Dirac's equation, as $g_0$ becomes infinitely large, $\delta_0$ becomes vanishingly small (i.e., much smaller than the Planck length $\lambda_P$).

In a five-dimensional Kaluza-Klein theory, which unifies gravity with electromagnetism, it can be shown that by unifying the electromagnetic field with gravity as dimensionless components of a Kaluza-Klein metric, both Newton's constant and the electric charge can be replaced in the Kaluza-Klein equations by the higher dimensional radius [22]. A reasonable goal for a Kaluza-Klein model is that all fields have the same physical dimensions, as well as that all physical constants should originate from geometry in a unified field theory. With this in mind, we can define a "geometric model" to be one in which all fields (like the gravitational field) are dimensionless, and with no physical constants (except mass) appearing explicitly in the Lagrangian. Defining higher dimensional bispinor fields on the Kaluza-Klein manifold [6], [13], [17], [18] does not satisfy these two conditions for a "geometric model". Recently, we showed that these conditions are uniquely satisfied in the tetrad formulation of the Kaluza-Klein model [22].

While it is generally agreed that the classical limit for (a large number) of photons is the classical electromagnetic field, it is also widely believed that no classical limit exists in the same sense for fermions [36], [42], [47]. However, the belief that fermions are not associated with a classical field is unfounded given that fermions, gravitons, and gauge bosons can be unified at a classical level as dimensionless components of a Kaluza-Klein metric together with scalar fields [20], [22]. In this formulation, the Kaluza-Klein Lagrangian precisely reduces to the usual Einstein-Maxwell-Dirac Lagrangian. Moreover, as shown in this paper, the unification at a classical level reveals the conditions for which Yang-Mills and Kaluza-Klein Lagrangians are equal.

## 2. Hestenes' tetrad and the tensor form of the Dirac Lagrangian

Because the Dirac gamma matrices are regarded as constant matrices, the Dirac equation, as described in most textbooks, is not covariant even under Lorentz transformations of the coordinates in the usual sense [31], [54]. A generally covariant tensor form of the Dirac Lagrangian was first derived by Takahashi using trace formulas of the Dirac matrices known as Fierz identities [33] – [35]. A simpler derivation, using trace formulas of the Pauli matrices, was presented as Appendix A and B of reference [55]. In this section we will give a straightforward derivation of the tensor form of the Dirac Lagrangian by using Hestenes' tetrad [32] as the reference tetrad for the spin connection [23] – [30]. For those familiar with spin connections this derivation will be the most direct. As in previous work, we show that the Dirac bispinor Lagrangian equals a constrained Yang-Mills Lagrangian in the limit of an infinitely large coupling constant. Both the constraint and the limit will be explicated in the Kaluza-Klein model in Section 3.

The existence of a globally defined tetrad of orthonormal vector fields is both necessary and sufficient for a noncompact four-dimensional space-time to admit a "spinor structure" [38] – [44]. At each point of the space-time, bispinors are defined relative to a reference tetrad of orthonormal vectors [23] – [30]. Usually in a Minkowski space-time the reference tetrad consists of coordinate vector fields associated with Cartesian coordinates, but this special choice of reference tetrad is not necessary.

In this section we will consider non-compact four-dimensional space-times with spinor structure [38] – [44]. A general reference tetrad, defined on an open subset U of the space-time, will be denoted by $e_a$ where $a = 0, 1, 2, 3$ is a tetrad index. Without further mentioning it, we will always assume that all bispinor and tetrad fields are smooth, except for some exceptional points where fields are allowed to be singular. We assume that these exceptional points are contained in the complement of an open dense subset of the space-time, which we define to be the complement of U. Also, without further mention, the space-time components of tensor fields will be defined for arbitrary coordinate charts in U. In each coordinate chart of U, we can express the tetrad $e_a$ as $e_a = e_a^\alpha \partial_\alpha$ where $\partial_\alpha$ denote the partial derivatives with respect to space-time coordinates $x^\alpha$ where $\alpha = 0, 1, 2, 3$, and $e_a^\alpha$ denote the tensor components of $e_a$. Tensor indices $\alpha, \beta, \gamma, \delta$ are lowered and raised using the space-time metric, denoted as $g_{\alpha\beta}$, and its inverse $g^{\alpha\beta}$. Tetrad indices a, b, c, d are lowered and raised using a Minkowski metric $g_{ab}$ (with diagonal elements $\{1, -1, -1, -1\}$ and zeros off the diagonal), and its inverse $g^{ab}$. Repeated tensor and tetrad indices will be summed from 0 to 3.

It has been shown that every bundle of spin frames is trivial on a non-compact four-dimensional space-time [40], [44]. Hence, on the open subset U of the space-time where the reference tetrad $e_a$ is defined, bispinor fields $\Psi$ can be simply expressed as maps $\Psi : U \to C^4$ with the spin group SL(2,C) acting on the complex vector space $C^4$. The canonical spin connection $\nabla_a$ acting on bispinor fields, induced from the Riemannian connection, is uniquely defined in U as follows [23] – [30].

$$\nabla_a = e_a^\alpha \partial_\alpha - \frac{i}{4} e_a^\alpha e_b^\beta (\nabla_\alpha e_{\beta c}) \sigma^{bc} \qquad (2.1)$$

where

$$\sigma^{bc} = \frac{i}{2}(\gamma^b \gamma^c - \gamma^c \gamma^b) \qquad (2.2)$$

and where $\nabla_\alpha$ denotes the Riemannian connection acting on the vector fields $e_a$ in each coordinate chart of U, and $\gamma^a$ are (constant) Dirac matrices [56]. By formula (2.1) the spin connection $\nabla_a$ acts on bispinor fields $\Psi$ defined on the open set U where the reference tetrad $e_a$ is defined. The reference tetrad $e_a$ transforms under the regular representation of the Lorentz group [23] – [31]. Note that the bispinor field $\Psi : U \to C^4$ has four complex scalar components that transform under the spinor representation of the Lorentz group whose Lie algebra generators are $-i\sigma_{ab}$ [24], [56]. The relationship between the spin connection $\nabla_a$ in formula (2.1) acting on bispinor fields $\Psi$ and the Riemannian connection $\nabla_\alpha$ acting on tensor fields will be further discussed below.

Dirac's bispinor Lagrangian in a Riemannian space-time is given by [24], [26], [27], [30]:

$$L_D = \text{Re}\,[i\overline{\Psi}\gamma^a\nabla_a\Psi - m_0 s] \tag{2.3}$$

where $m_0$ denotes the fermion mass and the complex scalar field s is defined by

$$\text{Re}\,[s] = \overline{\Psi}\Psi$$
$$\text{Im}\,[s] = i\overline{\Psi}\gamma^5\Psi \tag{2.4}$$

where (using bispinor notation) $\overline{\Psi} = \Psi^+\gamma^0$, and $\gamma^5 = i\gamma^0\gamma^1\gamma^2\gamma^3$ is the fifth Dirac matrix. (Throughout this section we adopt the bispinor notation of Bjorken and Drell [56]. In particular, for bispinors, $\Psi^*$ denotes the ordinary complex conjugate of $\Psi$, whereas, $\Psi^+$ denotes the Hermitian conjugate of $\Psi$). Formula (2.3) generalizes the usual Dirac bispinor Lagrangian for a Minkowski space-time, which uses the coordinate reference tetrad $e_a = \delta_a^\alpha \partial_\alpha$, where $\delta_a^\alpha$ equals one if $a = \alpha$ and zero otherwise. In Theorem 1, a different choice of reference tetrad $e_a$ (Hestenes' tetrad) will lead to the tensor form of the Dirac Lagrangian.

Except for the mass term, Dirac's bispinor Lagrangian (2.3) is invariant under $SL(2,R) \times U(1)$ gauge transformations acting on the bispinor field $\Psi$, with infinitesimal generators $\tau_K$ for $K = 0, 1, 2, 3$ defined by [56]:

$$\tau_0\Psi = -i\Psi, \qquad \tau_1\Psi = i\Psi^C$$
$$\tau_2\Psi = \Psi^C, \qquad \tau_3\Psi = i\gamma^5\Psi \tag{2.5}$$

where $\Psi^C = i\gamma^2\Psi^*$ denotes the charge conjugate of $\Psi$. Note that the action of $SL(2,R) \times U(1)$ on $\Psi$ is real linear, whereas usually only complex linear gauge transformations of bispinors are considered. The infinitesimal gauge generators $\tau_0, \tau_1, \tau_2$ generate $SL(2,R)$, and $\tau_3$ generates $U(1)$.

The $SL(2,R) \times U(1)$ gauge transformations generated by $\tau_K$ commute with Lorentz transformations [56]. From formula (2.5) the commutation relations of the gauge generators $\tau_K$ are given by [56]:

$$[\tau_0, \tau_1] = 2\tau_2, \quad [\tau_0, \tau_2] = -2\tau_1, \quad [\tau_1, \tau_2] = -2\tau_0 \tag{2.6}$$

and $\tau_3$ commutes with all the $\tau_K$. Formula (2.6) can be expressed as

$$[\tau_J, \tau_K] = 2 f_{JK}^L \tau_L \tag{2.7}$$

where $f_{JK}^L$ are the Lie algebra structure constants for the gauge group $SL(2,R) \times U(1)$. Note that from formula (2.6):

$$f_{JKL} = g_{LM} f_{JK}^M = -\varepsilon_{JKL3} \tag{2.8}$$

where $g_{LM}$ is the Minkowski metric (with diagonal elements $\{1, -1, -1, -1\}$ and zeros off the diagonal), and $\varepsilon_{JKLM}$ is the permutation tensor (with $\varepsilon_{0123} = -\varepsilon^{0123} = 1$). Gauge indices J, K, L, M are lowered and raised using the gauge metric $g_{JK}$, and its inverse $g^{JK}$. Repeated gauge indices are summed from 0 to 3.

The scalar field s in formula (2.4) is invariant under SL(2,R) gauge transformations, and transforms as a complex U(1) scalar under the U(1) gauge transformations (i.e., chiral gauge transformations [57]). To make the Lagrangian (2.3) invariant for all $SL(2,R) \times U(1)$ gauge transformations, it suffices that $m_0$ transform like $\bar{s}$ (the complex conjugate of s). Since $m_0$ appears in the Lagrangian (2.3) without derivatives, the assumption that $m_0$ transform like $\bar{s}$ under U(1) chiral gauge transformations, has no effect on the Dirac equation. For example, an alternative form of the four dimensional Dirac equation, proposed by Dirac in 1935, rotates the bispinor field $\Psi$ by the U(1) chiral gauge transformation $e^{-(i\pi/4)\gamma^5}$ or, equivalently, the scalar field s by $e^{-i\pi/2}$ so that $m_0$ is replaced by $im_0$ in the bispinor Lagrangian [58], [59]. The mass term in the bispinor Lagrangian then becomes $im_0 \bar{\Psi} \gamma^5 \Psi$ instead of $-m_0 \bar{\Psi} \Psi$ (see formulas (2.3) and (2.4)).

From the Dirac bispinor Lagrangian (2.3) we can derive the following $SL(2,R) \times U(1)$ Noether currents:

$$j^K = e^a \operatorname{Re}[i \bar{\Psi} \gamma_a \tau^K \Psi] \tag{2.9}$$

Specifically, substituting the infinitesimal gauge generators $\tau_K$ from formula (2.5) into (2.9) we have

$$j_a^0 = \bar{\Psi} \gamma_a \Psi, \qquad j_a^1 = \operatorname{Re}[\bar{\Psi} \gamma_a \Psi^C]$$

$$j_a^2 = \operatorname{Im}[\bar{\Psi} \gamma_a \Psi^C], \qquad j_a^3 = \bar{\Psi} \gamma_a \gamma^5 \Psi \tag{2.10}$$

where $j_a^K$ denote the tetrad components of $j^K = j_a^K e^a$. Note that $j^0$, $j^1$, and $j^2$ are SL(2,R) Noether currents and $j^3$ is the U(1) Noether current. In particular $j^0$ is the electromagnetic current and $j^3$ is the chiral current. The real Noether currents $j^K$ and complex scalar field s satisfy an orthogonal constraint known as a Fierz identity [33], [34], [55]:

$$j_a^K j_{Kb} = |s|^2 g_{ab}$$

$$j_a^J j^{Ka} = |s|^2 g^{JK}$$

(2.11)

Note that since the infinitesimal gauge generators $\tau_K$ and the Dirac matrices $\gamma^a$ are constants, both the scalar field s and the tetrad components of the Noether currents $j_a^K$ depend only on the bispinor field $\Psi$. That is from formulas (2.4) and (2.10) both $j_a^K$ and s are bilinear functions of the components of $\Psi$. Using the Leibniz product rule for covariant derivatives, the spin connection $\nabla_a$ defined in formula (2.1) acts on $j_a^K$ and s by its action on $\Psi$. Using the fact that the infinitesimal gauge generators $\tau_K$ commute with the infinitesimal Lorentz transformations $-i\sigma_{ab}$, and using the bispinor identities given in formula (2.23), it is easily shown that the spin connection $\nabla_a$ acting on $j_a^K$ and s is related to the Riemannian connection $\nabla_\alpha$ by

$$\nabla_a j_b^K = (\nabla_\alpha j_\beta^K) e_a^\alpha e_b^\beta$$

$$\nabla_a s = (\nabla_\alpha s) e_a^\alpha$$

(2.12)

where $j_\beta^K = j_b^K e_\beta^b$. This relationship justifies the definition of the canonical spin connection $\nabla_a$ in formula (2.1).

A derivation of the tensor form of Dirac's bispinor Lagrangian (2.3) follows from the map $\Psi \to (j_a^K, s)$. Apart from the singular set where s vanishes, we can make a special choice of orthonormal reference tetrad as follows:

$$e_a = |s|^{-1} \delta_a^K j_K$$

(2.13)

where $\delta_a^K$ equals one if $a = K$ and zero otherwise. The following lemma shows that relative to this special reference tetrad, which is Hestenes' tetrad [32], the bispinor field

Ψ at each point in the space-time is "at rest", and Ψ becomes locally a function of a complex scalar field σ, which has s as its square.

**LEMMA:** Relative to Hestenes' tetrad (2.13), at each space-time point where Hestenes' tetrad is defined, every bispinor field Ψ has the form:

$$\Psi = \begin{bmatrix} 0 \\ \text{Re}[\sigma] \\ 0 \\ -i\,\text{Im}[\sigma] \end{bmatrix} \qquad (2.14)$$

where σ is a locally defined complex scalar field, which has s as its square.

**PROOF:** Given $j^K$ and s, we will solve for Ψ. Substituting $j^K$ defined by formula (2.9) into formula (2.13) for Hestenes' tetrad, gives

$$j_a^K = \text{Re}[i\overline{\Psi}\gamma_a \tau^K \Psi] = |s|\delta_a^K \qquad (2.15)$$

It is then straightforward to verify that all solutions of equations (2.4) and (2.15) are of the form (2.14) with the complex scalar σ having s as its square. **Q. E. D.**

Note that choosing Hestenes' tetrad as the reference tetrad reduces the bispinor field Ψ to locally depend only on a scalar field σ, at all points where Hestenes' tetrad is defined. Substitution of formula (2.14) for Ψ into formula (2.3), expresses the Dirac bispinor Lagrangian in terms of Hestenes' tensor fields $(e_a, \sigma)$. Further examination of formulas (2.1), (2.3), and (2.14) shows that the Dirac Lagrangian can be expressed solely with the tensor fields $(j^K, s)$. This result is proved below in Theorem 1, and was first derived by Takahashi, who used Fierz identities [33], [34].

To show that the tensor form of Dirac's bispinor Lagrangian (2.3) is a constrained Yang-Mills Lagrangian in the limit of an infinitely large coupling constant, we map SL(2,R) × U(1) gauge potentials $F_\alpha^K$ and a complex scalar field ρ into $(j^K, s)$ by setting:

$$j_\alpha^K = 4|\rho|^2 F_\alpha^K$$

$$s = 4|\rho|^2 \overline{\rho} \qquad (2.16)$$

where $j^K_\alpha = j^K_a e^a_K$ are the tensor components of $j^K$. From formulas (2.11) and (2.16), since the reference tetrad $e_a$ is orthonormal, the gauge potentials $F^K_\alpha$ satisfy an orthogonal constraint, which can be expressed in two equivalent ways:

$$F^K_\alpha F_{K\beta} = |\rho|^2 g_{\alpha\beta}$$

$$F^J_\alpha F^{K\alpha} = |\rho|^2 g^{JK}$$
(2.17)

Consider the following Yang-Mills Lagrangian for the gauge potentials $F^K_\alpha$ and the complex scalar field $\rho$:

$$L_g = \frac{1}{4g} F^K_{\alpha\beta} F^{\alpha\beta}_K + \frac{1}{g_0} \overline{D_\alpha(\rho+\mu)} D^\alpha(\rho+\mu)$$
(2.18)

where, because of the symmetry of the Riemannian connection, the Yang-Mills field tensor $F^K_{\alpha\beta}$ is given by

$$F^K_{\alpha\beta} = \nabla_\alpha F^K_\beta - \nabla_\beta F^K_\alpha + g f^K_{MN} F^M_\alpha F^N_\beta$$
(2.19)

and where the Yang-Mills coupling constant $g$ is a self-coupling of the gauge potentials $F^K_\alpha$. Furthermore, in the Lagrangian (2.18), the complex Higgs scalar $\mu$ satisfies [57]:

$$\mu = \frac{2m_0}{g_0}, \qquad \partial_\alpha \mu = 0$$
(2.20)

where $m_0$ is the fermion mass, and $g_0 = \frac{3}{2} g$. As previously stated for Dirac's bispinor Lagrangian (2.3) both the complex scalar field s and the fermion mass $m_0$ transform as U(1) scalars. The same is true for $\rho$ and $\mu$ by formulas (2.16) and (2.20). Hence the covariant derivative $D_\alpha$ acts on $\rho + \mu$ as follows:

$$D_\alpha(\rho+\mu) = \partial_\alpha \rho - ig_0 F^3_\alpha(\rho+\mu)$$
(2.21)

That is, $g_0$ is the Yang-Mills constant which couples the U(1) scalars $\rho$ and $\mu$ to the U(1) gauge potential $F^3_\alpha$. Note that from formula (2.8), the Lie algebra structure constants $f^L_{JK}$ vanish if any gauge index J, K, L equals 3, so that $g_0$ can be different than $g$, and we have defined $g_0 = \frac{3}{2} g$.

**THEOREM 1:** At every space-time point where Hestenes' tetrad is defined, Dirac's bispinor Lagrangian (2.3) equals the Yang-Mills Lagrangian (2.18) in the limit of a large coupling constant. That is,

$$L_D = \lim_{g \to \infty} L_g \tag{2.22}$$

**PROOF:** From formulas (2.1), (2.2), (2.3), (2.10), and using the following identities for Dirac matrices [60]:

$$\gamma^a \gamma^b \gamma^c = g^{ab}\gamma^c - g^{ac}\gamma^b + g^{bc}\gamma^a + i\varepsilon^{abcd}\gamma^5 \gamma_d$$

$$\gamma^a \gamma^5 = -\gamma^5 \gamma^a \tag{2.23}$$

$$(\gamma^a)^+ \gamma^0 = \gamma^0 \gamma^a$$

we can express Dirac's bispinor Lagrangian (2.3) in a Riemannian space-time as a sum of two terms:

$$L_D = \text{Re}[i\overline{\Psi}\gamma^a e_a^\alpha \partial_\alpha \Psi - m_0 s] + \frac{1}{4}\varepsilon^{abcd} e_a^\alpha e_b^\beta (\nabla_\alpha e_{\beta c}) j_d^3 \tag{2.24}$$

We will express each of these terms with the tensor fields $(F_\alpha^K, \rho)$. Substituting formula (2.14) for $\Psi$ in the first term of $L_D$, we get using formula (2.16),

$$\text{Re}[i\overline{\Psi}\gamma^a e_a^\alpha \partial_\alpha \Psi - m_0 s] = \text{Re}[2i\overline{\rho} F_\alpha^3 \partial^\alpha \rho - 4m_0 |\rho|^2 \overline{\rho}] \tag{2.25}$$

Noting that $j_d^3 = |s|\delta_d^3$ by formulas (2.9) and (2.15), and using formulas (2.13) and (2.16), the second term of $L_D$ becomes:

$$\frac{1}{4}\varepsilon^{abcd} e_a^\alpha e_b^\beta (\nabla_\alpha e_{\beta c}) j_d^3 = -(\nabla_\alpha \mathbf{F}_\beta) \bullet \mathbf{F}^\alpha \times \mathbf{F}^\beta \tag{2.26}$$

where $\mathbf{F}_\alpha = (F_\alpha^0, F_\alpha^1, F_\alpha^2)$. Summing the two terms (2.25) and (2.26), formula (2.24) becomes:

$$L_D = -(\nabla_\alpha \mathbf{F}_\beta) \bullet \mathbf{F}^\alpha \times \mathbf{F}^\beta + \text{Re}[2i\overline{\rho} F_\alpha^3 (\nabla^\alpha \rho) - 4m_0 |\rho|^2 \overline{\rho}] \tag{2.27}$$

To complete the proof it remains to show that formula (2.27) equals the limit in (2.22). Observe that all terms in the Yang-Mills Lagrangian (2.18) which are quartic in the fields $(F_\alpha^K, \rho)$ cancel by virtue of the orthogonal constraint (2.17) and the relation between the coupling constants $g_0 = (3/2)g$. Quadratic terms in the fields $(F_\alpha^K, \rho)$ vanish in the limit (2.22). Thus, the limit (2.22) only contains terms cubic in the fields $(F_\alpha^K, \rho)$. The cubic terms of the Yang-Mills Lagrangian (2.18) are given by:

$$L_g^{(3)} = f_{JKL}(\nabla_\alpha F_\beta^J)F^{K\alpha}F^{L\beta} + \text{Re}[2i\bar{\rho}F_\alpha^3(\nabla^\alpha \rho) + 4m_0 F_\alpha^3 F^{3\alpha}\bar{\rho}] \qquad (2.28)$$

which equals $L_D$ given in formula (2.27), after applying the orthogonal constraint (2.17) to obtain

$$F_\alpha^3 F^{3\alpha} = -|\rho|^2 \qquad (2.29)$$

and using formula (2.8) to replace the triple vector product with the Lie algebra structure constants $f_{JKL}$. **Q. E. D.**

Since Theorem 1 shows that the Dirac bispinor Lagrangian (2.3) and its tensor form (2.22) are equal at all space-time points where Hestenes' tetrad is defined, we will briefly discuss the physical interpretation of the singularities, where Hestenes' tetrad is not defined. By formula (2.13) Hestenes' tetrad $e_a$ is defined wherever the scalar field s does not vanish. When s vanishes there are two types of singularities. First, if the bispinor field $\Psi$ vanishes, both s and its first partial derivatives vanish by formula (2.4), and the tensor form of the Dirac equation allows $e_a$ to be arbitrary. At such space-time points the tensor fields $F_\alpha^K$ and $\rho$ in the Lagrangian (2.18) vanish. Second, if s vanishes but $\Psi$ does not, then the nonvanishing fermion particle current lies on the light cone [32]. For physical solutions representing massive fermions, these singularities must form an exceptional (nowhere dense) set. Thus singularities in the tensor fields $F_\alpha^K$ and $\rho$ can only occur in the complement of an open dense subset of the space-time. Consequently, putative differences between the bispinor field $\Psi$ and the tensor fields $F_\alpha^K$ and $\rho$ cannot be observed in experiments.

Note also from the Lagrangian (2.18) that we can derive all bispinor observables (e.g., the energy-momentum tensor $T^{\alpha\beta}$, spin polarization tensor $S^{\alpha\beta\gamma}$, and the electric current vector $J^\alpha$) directly from the Yang-Mills formulas. For example, the Dirac spin polarization tensor $S^{\alpha\beta\gamma}$ is usually expressed as follows using bispinor notation [61]:

$$S^{\alpha\beta\gamma} = -\frac{1}{4}\bar{\Psi}(\gamma^\alpha \sigma^{\beta\gamma} + \sigma^{\beta\gamma}\gamma^\alpha)\Psi \qquad (2.30)$$

where $\gamma^\alpha = e_a^\alpha \gamma^a$ and where $\sigma^{\alpha\beta} = (i/2)(\gamma^\alpha\gamma^\beta - \gamma^\beta\gamma^\alpha)$. Using the following identity [60]:

$$\gamma^\alpha \sigma^{\beta\gamma} + \sigma^{\beta\gamma} \gamma^\alpha = 2\varepsilon^{\alpha\beta\gamma\delta} \gamma_\delta \gamma^5 \qquad (2.31)$$

together with formulas (2.10), (2.16), and (2.17), formula (2.30) reduces to:

$$S_{\alpha\beta\gamma} = -\frac{1}{2}\varepsilon_{\alpha\beta\gamma\delta} \overline{\Psi}\gamma^\delta\gamma^5\Psi = -\frac{1}{2}\varepsilon_{\alpha\beta\gamma\delta} j^{3\delta} = 2\,\mathbf{F}_\alpha \bullet \mathbf{F}_\beta \times \mathbf{F}_\gamma \qquad (2.32)$$

The Yang-Mills version of the spin polarization tensor is easily shown from the Lagrangian (2.18) to be [61], [62]:

$$S_g^{\alpha\beta\gamma} = \frac{1}{g}\text{Re}[F_K^{\alpha\beta} F^{K\gamma} - F_K^{\alpha\gamma} F^{K\beta}] \qquad (2.33)$$

In the limit of a large coupling constant g, the Yang-Mills formula (2.33) becomes using the definition of $F_K^{\alpha\beta}$ given in formula (2.19):

$$\lim_{g\to\infty} S_g^{\alpha\beta\gamma} = 2\,\mathbf{F}^\alpha \bullet \mathbf{F}^\beta \times \mathbf{F}^\gamma \qquad (2.34)$$

which equals $S^{\alpha\beta\gamma}$ by formula (2.32). Similarly, we can derive $T^{\alpha\beta}$ and $J^\alpha$ directly from the Yang-Mills formulas.

We mention in passing that, just as for Yang-Mills fields, the bispinor canonical (non-symmetric) energy-momentum tensor $T^{\alpha\beta}$ and spin polarization tensor $S^{\alpha\beta\gamma}$ satisfy a relation [62]:

$$\partial_\alpha S^{\alpha\beta\gamma} - T^{\beta\gamma} + T^{\gamma\beta} = 0 \qquad (2.35)$$

From this relation we can define a symmetric energy-momentum tensor, which is also conserved as follows:

$$\Theta^{\alpha\beta} = T^{\alpha\beta} + \frac{1}{2}\partial_\gamma (S^{\beta\gamma\alpha} + S^{\alpha\gamma\beta} - S^{\gamma\alpha\beta}) \qquad (2.36)$$

In general relativity, the symmetric tensor $\Theta^{\alpha\beta}$ is the bispinor source of the gravitational field, which is derived by varying the action with respect to the metric tensor [62]. (The action is formed of the Lagrangian (2.18) with the orthogonal constraint (2.17) expressed using Lagrange multiplyers.) Note that the general relativistic derivation of a symmetric energy-momentum tensor $\Theta^{\alpha\beta}$ is more self-evident using the Yang-Mills formulas rather than the bispinor formulas [30]. Also, for those interested in torsion theory generalizations, the interaction with torsion is much simpler to derive using the Yang-Mills formulas [29].

## 3. Conditions for equality of the Einstein-Dirac and Kaluza-Klein Lagrangians

In this section we will derive the conditions for equality of the Einstein-Dirac Lagrangian and the Lagrangian in the Kaluza-Klein tetrad model. This model explicates both the orthogonal constraint (2.17) and the limit (2.22). It will be shown that while the orthogonal constraint is inherent in the structure of the tetrads, the limit implies that the radius of the higher compact dimensions is vanishingly small compared with the Planck length. We will show that equality of the Einstein-Dirac and Kaluza-Klein Lagrangians requires restricting to horizontal vector fields and horizontal sectional curvatures.

Let $M = X \times G$ be the Kaluza-Klein manifold, with X a four dimensional space-time, and G the four-dimensional real Lie group $SL(2,R) \times U(1)$. On the space-time X, let $\beta^K$ be a global, nonsingular tetrad of one-forms with K = 0, 1, 2, 3. (This assumption, which is equivalent to the existence of spinor structure, simplifies the derivations in this section, but is stronger than necessary. As discussed in Section 2, the tetrad field $\beta^K$ is allowed to be singular at exceptional points in X.) The gravitational field on X, which we denote as $\beta$, is defined to be the unique metric tensor with the Minkowski signature, for which the tetrad $\beta^K$ is orthonormal. That is:

$$\beta = g_{JK} \beta^J \otimes \beta^K \qquad (3.1)$$

where

$$g_{JK} = g^{JK} = \begin{bmatrix} 1 & 0 & 0 & 0 \\ 0 & -1 & 0 & 0 \\ 0 & 0 & -1 & 0 \\ 0 & 0 & 0 & -1 \end{bmatrix} \qquad (3.2)$$

The tetrad of smooth one-forms $\beta^K$ uniquely determines its dual tetrad of smooth vector fields $b_K$ on X satisfying

$$\beta^K(b_J) = \delta_J^K \qquad (3.3)$$

where $\delta_J^K$ equals one if J = K, and equals zero otherwise. From formula (3.1), the vector fields $b_K$ form an orthonormal basis for each tangent space of X.

The fermion field on X we denote as $(F^K, \rho)$, where $\rho$ is a complex scalar field and $F^K = |\rho| \beta^K$. Thus the dynamical fields are the tetrad of one forms $\beta^K$ and $\rho$. We will show that the gravitational field $\beta$ and the bispinor field $\Psi$ (which together have $10 + 8 = 18$ real components), are represented faithfully by $\beta^K$ and $\rho$ (which also have $16 + 2 = 18$ real components). Note that this parsimonious feature has not been achieved in previous theories of fermions in a curved space-time, in which the tetrad field adds six supernumerary boson fields to the gravitational and bispinor fields [6], [13], [17], [18]. We will then derive the usual Einstein-Dirac Lagrangian from the Kaluza-Klein Lagrangian (3.22) for the fields $\beta^K$ and $\rho$.

On G, the four-dimensional real Lie group $SL(2,R) \times U(1)$, we fix a nonsingular tetrad of right-invariant one-forms $\alpha^K$ with K = 0, 1, 2, 3. The tetrad of right-invariant one-forms $\alpha^K$ defines a right-invariant metric on the Lie group G given by:

$$\alpha = g_{JK} \alpha^J \otimes \alpha^K \tag{3.4}$$

where $g_{JK}$ has the same form as the Minkowski metric in the definition (3.2). Since G is a four-dimensional Lie group, the $\alpha^K$ form a basis for the dual of the Lie algebra of G.

For vector fields v and w on G, we will denote the inner product with respect to the metric $\alpha$ by $\langle v, w \rangle$, that is:

$$\langle v, w \rangle = \alpha(v, w) = g_{JK} \alpha^J(v) \alpha^K(w) \tag{3.5}$$

The tetrad of right-invariant one-forms $\alpha^K$ uniquely determines a dual tetrad of right-invariant vector fields $a_K$ on G satisfying

$$\alpha^K(a_J) = \delta^K_J \tag{3.6}$$

The right-invariant vector fields $a_K$ form a basis for the Lie algebra of G. This basis is orthonormal, since from formulas (3.5) and (3.6) we get:

$$\langle a_J, a_K \rangle = g_{JK} \tag{3.7}$$

We can choose the fixed tetrad $\alpha^K$ so that the vector fields $a_K$ satisfy the following $SL(2,R) \times U(1)$ commutation relations:

$$[a_0, a_1] = \delta^{-1} a_2$$

$$[a_0, a_2] = -\delta^{-1} a_1 \qquad (3.8)$$

$$[a_1, a_2] = -\delta^{-1} a_0$$

where $\delta$ is a length parameter. All other commutators vanish. As usual in general relativity, both length and time carry the same unit. As on any physical manifold, the one-forms $\alpha^K$ carry units of length, so that their duals, the vector fields $a_K$ in formula (3.8), carry units of mass (i.e., inverse length). From formulas (3.7) and (3.8) it is evident that $\delta$ is the radius of the U(1) subgroups of SL(2,R). Formula (3.8) can be written more succinctly as:

$$[a_J, a_K] = \frac{1}{\delta} f_{JK}^L a_L \qquad (3.9)$$

which defines the Lie algebra structure constants $f_{JK}^L$. Note that the structure constants $f_{JK}^L$ are dimensionless, so that the length parameter $\delta$ is required in formula (3.9) to balance the dimensions. Also, in formula (3.4), the metric constants $g_{JK}$ are dimensionless. Although we do not make use of the following property in the tetrad Kaluza-Klein model, note from formulas (3.2) and (3.8) that $f_{JKL} = g_{LM} f_{JK}^M$ is completely antisymmetric in the indices J, K, and L. When this property holds, the metric is called "bi-invariant", since it is both right and left invariant [63]. We will see generally that the tetrad Kaluza-Klein model does not require that the right-invariant metric $\alpha$ given in formula (3.4) be bi-invariant.

Note that while the orthonormal and commutation relations (3.7) and (3.8) determine the radius of the U(1) subgroups of SL(2,R), they do not determine the radius of the U(1) factor of the Lie group G = SL(2,R) × U(1). The radius of the U(1) factor of G will be denoted as $\delta_0$. The ratio $\delta/\delta_0$ is a parameter which we can equate to the ratio $g_0/g$ of coupling constants in the Yang-Mills Lagrangian (2.18). That is, the length parameters $\delta_0$ and $\delta$ of the tetrad Kaluza-Klein model will be set as $\delta_0 = (2/3)\delta$ in correspondence with $g_0 = (3/2)g$ in the Lagrangian (2.18).

Thus on the Kaluza-Klein manifold M = X × G, we can define a fixed tetrad of one-forms $\alpha^K$ and a dynamic tetrad of one-forms $\beta^K$ induced from the projections of M onto its factors G and X. ($\alpha^K$ and $\beta^K$ on M are the pullbacks of $\alpha^K$ on G and $\beta^K$ on X by the projection maps.) We define a third tetrad of one-forms $v^K$ on M by:

$$v^K = \alpha^K - (\kappa\delta)^{1/3}|\rho|\beta^K \qquad (3.10)$$

where $\kappa$ is $16\pi/3$ times Newton's constant $\kappa_0$, and $\rho$ is a complex scalar field on M. Note that the constant $\kappa$ has dimension of length squared, the constant $\delta$ has dimension of length as in formula (3.9), the scalar field $\rho$ has dimension of mass, and the one-forms $\alpha^K$, $\beta^K$, and $v^K$ each have dimension of length.

The one-forms $(\beta^K, v^K)$ form a basis for each cotangent space of $M = X \times G$. The Kaluza-Klein metric on M is defined to be:

$$\gamma = g_{JK}(\beta^J \otimes \beta^K + v^J \otimes v^K) \qquad (3.11)$$

which depends only on the dynamical fields $\beta^K$ and $\rho$, since $\alpha^K$ in formula (3.10) is fixed by the basis chosen for the Lie algebra of G.

To demonstrate that $\gamma$ is a Kaluza-Klein metric, we define local coordinate one-forms $dx^\alpha$ with $\alpha = 0,1,2,3$ on a local coordinate chart of X. The gravitational field $\beta$ is expressed locally by:

$$\beta = g_{\alpha\beta}\, dx^\alpha \otimes dx^\beta \qquad (3.12)$$

Writing $\beta^K = \beta^K_\alpha\, dx^\alpha$, we obtain from formulas (3.1) and (3.12):

$$g_{\alpha\beta} = g_{JK}\, \beta^J_\alpha \beta^K_\beta \qquad (3.13)$$

If we choose $(dx^\alpha, \alpha^K)$ for a basis of one-forms, then from formulas (3.10) and (3.13), the Kaluza-Klein metric (3.11) has the following components:

$$\gamma = \begin{bmatrix} g_{\alpha\beta} + \lambda^2 g_{JK} F^J_\alpha F^K_\beta & -\lambda F^J_\alpha g_{JK} \\ -\lambda g_{JK} F^K_\beta & g_{JK} \end{bmatrix} \qquad (3.14)$$

where $\lambda = (\kappa\delta)^{1/3}$ is a Kaluza-Klein parameter having dimension of length [9], and

$$F^K_\alpha = |\rho|\beta^K_\alpha \qquad (3.15)$$

Thus, $\gamma$ is precisely the Kaluza-Klein metric [9] for the gravitational field $g_{\alpha\beta}$ and the gauge potentials $F^K_\alpha$. By formulas (3.13) and (3.15) the $F^K_\alpha$ satisfy:

$$g_{JK} F_\alpha^J F_\beta^K = |\rho|^2 g_{\alpha\beta} \qquad (3.16)$$

which is precisely the orthogonal constraint (2.17). Furthermore, by formula (3.13), the gravitational field $g_{\alpha\beta}$ has the same (Minkowski) signature as $g_{JK}$ on G.

We denote the vector fields dual to $(\beta^K, \alpha^K)$ as $(b_K, a_K)$. The vector fields dual to $(\beta^K, v^K)$ are then the horizontal and vertical vector fields $(v_K, a_K)$, where from formula (3.10):

$$v_K = b_K + (\kappa\delta)^{1/3} |\rho| a_K \qquad (3.17)$$

From formula (3.11), the vector fields $(v_K, a_K)$ form an orthonormal basis with respect to the Kaluza-Klein metric $\gamma$ on each tangent space of M.

We extend the inner product notation in formula (3.5) to vector fields v and w defined on M as follows:

$$\langle v, w \rangle = \gamma(v, w) = g_{JK}[\beta^J(v)\beta^K(w) + v^J(v)v^K(w)] \qquad (3.18)$$

Thus, for the orthonormal vector fields $v_K$ and $a_K$ defined on M:

$$\langle v_J, v_K \rangle = \langle a_J, a_K \rangle = g_{JK}$$

$$\langle v_J, a_K \rangle = 0 \qquad (3.19)$$

for all indices J, K = 0, 1, 2, 3. That is, with respect to the basis $(v_K, a_K)$, the Kaluza-Klein metric $\gamma$ becomes:

$$\gamma = \begin{bmatrix} g_{JK} & 0 \\ 0 & g_{JK} \end{bmatrix} \qquad (3.20)$$

The manifold $M = X \times G$ has a natural right action of G defined by $h(x, g) = (x, gh)$ for each $(x, g) \in M$ and $h \in G$. For $v_K$ to be right invariant, it is necessary and sufficient that $b_K$ and $|\rho|$ depend only on the space-time coordinates $x \in X$. For Kaluza-Klein solutions of interest, the complex scalar field $\rho$ has the form: (See discussion of $\rho$ following Theorem 2.)

$$\rho = e^{iy/\delta_0} \tilde{\rho}(x) \tag{3.21}$$

where y is a global coordinate of the U(1) factor of the gauge group SL(2,R) × U(1) for which $a_3 = -\partial/\partial y$ is a unit vector field.

Our goal in this section is to derive the Einstein and Dirac Lagrangians from the following Lagrangian for the fields $(\beta^K, \rho)$:

$$L = \frac{1}{16\pi\kappa_0} R_v + \frac{1}{g_0} \overline{v_K(\rho+\mu)} \, v^K(\rho+\mu) \tag{3.22}$$

where $\kappa_0$ and $g_0$ are constants ($\kappa_0$ is Newton's constant), and where $v^K = g^{JK} v_J$. The mass parameter $\mu$ is defined on M by:

$$\mu = e^{iy/\delta_0} \tilde{\mu} \tag{3.23}$$

where $\tilde{\mu}$ is a constant. ($\mu$ acts as a Higgs field in this model. See discussion of the scalar fields $\rho$ and $\mu$ following Theorem 2.) $R_v$ is the sum of sectional curvatures over the four-dimensional horizontal subspaces spanned by the orthonormal tetrad $v_K$ in each tangent space of M:

$$R_v = g^{JK} g^{LM} \langle R(v_J, v_L) v_K, v_M \rangle \tag{3.24}$$

where R( , ) is the curvature two-form [63] associated with the Kaluza-Klein metric $\gamma$ on M.

Let $d\gamma$ denote the volume form on M = X × G defined by the Kaluza-Klein metric $\gamma$. (We do not confuse the symbol "d" with exterior differentiation since the metric $\gamma$ is not a differential form.) Similarly let $d\alpha$ and $d\beta$ denote the volume forms defined by the metrics $\alpha$ and $\beta$ on the manifolds G and X, respectively. Note that $d\alpha$ is a fixed volume form on G, whereas $d\beta$ depends on the dynamic fields $\beta^K$. Since the one-forms $(\beta^K, v^K)$ are orthonormal, we see from formula (3.10) that

$$d\gamma = d\beta \wedge d\alpha \tag{3.25}$$

Therefore, the action associated with the Lagrangian (3.22) is given by:

$$S = \int L(\beta^K, \rho) \, d\beta \wedge d\alpha \tag{3.26}$$

Note that in the action (3.26), the gravitational field $g_{\alpha\beta}$ and the bispinor field $\Psi$, which together have $10 + 8 = 18$ real components, are represented by $\beta^K$ and $\rho$, which also have $16 + 2 = 18$ real components [19] – [22]. We show in the following theorem that the Lagrangian (3.22) equals the Hilbert-Einstein Lagrangian for the gravitational field plus the Dirac-Yang-Mills Lagrangian (2.18). The orthogonal constraint (2.17) of the Dirac-Yang-Mills equation has already been shown to be a consequence of the tetrad in formula (3.16).

**THEOREM 2:** If we define the constants $\kappa$, $g$, $g_0$ in terms of Newton's constant $\kappa_0$ and the length parameters $\delta$ and $\delta_0$ as follows:

$$\kappa = \frac{16\pi}{3}\kappa_0$$

$$g = \frac{(\kappa\delta)^{1/3}}{\delta} \tag{3.27}$$

$$g_0 = \frac{(\kappa\delta)^{1/3}}{\delta_0}$$

then the total Lagrangian L given in formula (3.22) equals the Hilbert-Einstein Lagrangian $(16\pi\kappa_0)^{-1} R_X$ for the gravitational field plus the Dirac-Yang-Mills Lagrangian $L_g$ given in formula (2.18) for the fermion field, and similarly for the total action (3.26). Furthermore, the limit (2.22) required to obtain Dirac's bispinor equation, forces the length parameters $\delta$ and $\delta_0$ in the Kaluza-Klein model to become vanishingly small compared with the Planck length $\lambda_P = \kappa_0^{1/2}$.

**PROOF:** We will derive an alternative local expression for the Lagrangian (3.22), which simplifies the computations. Define a local coordinate tetrad $v_\alpha$ as follows:

$$v_\alpha = \partial_\alpha + (\kappa\delta)^{1/3} F_\alpha^K a_K \tag{3.28}$$

where $\partial_\alpha$ are the coordinate vector fields dual to $dx^\alpha$ in formula (3.12). Since $v_\alpha = \beta_\alpha^K v_K$, the tetrads $v_K$ and $v_\alpha$ in formulas (3.17) and (3.28) span the same four dimensional horizontal distribution over the Kaluza-Klein manifold M.

The inverse relation, $v_K = b_K^\alpha v_\alpha$, where $b_K^\alpha$ are the components of the vector fields $b_K = b_K^\alpha \partial_\alpha$, follows from formulas (3.3), (3.15), (3.17), and (3.28). Similarly, formulas (3.3) and (3.13) imply:

$$b_K^\beta = g_{JK} g^{\alpha\beta} \beta_\alpha^J$$

$$g^{\alpha\beta} = g^{JK} b_J^\alpha b_K^\beta$$

(3.29)

Then, substituting $v_K = b_K^\alpha v_\alpha$ into $R_v$, the sum of sectional curvatures over the horizontal distribution spanned by $v_K$ in formula (3.24), gives:

$$R_v = g^{\alpha\beta} g^{\gamma\delta} \langle R(v_\alpha, v_\gamma) v_\beta, v_\delta \rangle \qquad (3.30)$$

and the Lagrangian (3.22) equals:

$$L = \frac{1}{16\pi\kappa_0} R_v + \frac{1}{g_0} \overline{v_\alpha(\rho+\mu)} \, v^\alpha(\rho+\mu) \qquad (3.31)$$

Formula (3.30) is evaluated by computing $R_v$ using the vector fields $(v_\alpha, a_K)$ as a basis on M. Note that with respect to this basis, the Kaluza-Klein metric (3.11) has the following components:

$$\gamma = \begin{bmatrix} g_{\alpha\beta} & 0 \\ 0 & g_{JK} \end{bmatrix} \qquad (3.32)$$

The local expressions of $v_\alpha$, $R_v$, and $\gamma$ given in formulas (3.28), (3.30), and (3.32) are equal to the usual expressions in Kaluza-Klein theory [9]. A straightforward derivation using the commutation relations (3.9) shows that:

$$R_v = R_X + \frac{3}{4}(\kappa\delta)^{2/3} F_{\alpha\beta}^K F_K^{\alpha\beta} \qquad (3.33)$$

where $R_X$ denotes the scalar curvature of X, and

$$F_{\alpha\beta}^K = \partial_\alpha F_\beta^K - \partial_\beta F_\alpha^K + g f_{MN}^K F_\alpha^M F_\beta^N \qquad (3.34)$$

where $g = (\kappa/\delta^2)^{1/3}$. Note in formula (3.33) that indices are raised and lowered in the obvious way. That is

$$F_K^{\alpha\beta} = g^{\gamma\alpha} g^{\delta\beta} g_{JK} F_{\gamma\delta}^J \qquad (3.35)$$

(Because in formula (3.24), we restricted $R_v$ to the tetrad $v_K$, the scalar curvature of G does not occur in formula (3.33)).

Having computed $R_v$ in formula (3.33), and choosing the constants $\kappa$, $g$, and $g_0$ as in formula (3.27), we see in formula (3.31) that the total Lagrangian L equals the Hilbert-Einstein Lagrangian for $g_{\alpha\beta}$ plus the Dirac-Yang-Mills Lagrangian $L_g$ given in formula (2.18) for $F_\alpha^K$ and $\rho$.

Furthermore, since $\delta = (3/2)\delta_0$, formula (3.27) gives:

$$\delta_0 = \left(\frac{8\pi}{g_0^3}\right)^{1/2} \lambda_P \qquad (3.36)$$

which relates the Kaluza-Klein radius $\delta_0$ to the Planck length $\lambda_P = \kappa_0^{1/2}$. Thus in the limit required to obtain Dirac's equation, that is as $g_0$ becomes infinitely large, $\delta_0$ must become vanishingly small compared to the Planck length. The same is true for the radius $\delta = (3/2)\delta_0$. **Q. E. D.**

The Lagrangian (3.22) generalizes as follows. From formulas (3.10), (3.11), (3.13), and (3.15), and from the local expressions of the Lagrangian in formulas (3.28), (3.30), and (3.31), the gauge group G of the Kaluza-Klein manifold $M = X \times G$ can be generalized to larger Lie groups of dimension d > 4. For such generalizations we define $v_K = b_K^\beta v_\beta$ where $b_K^\beta = g^{\alpha\beta} g_{JK} \beta_\alpha^J$. Although, the d global vector fields $v_K$ are too many to form a tetrad when d > 4, they span a four dimensional horizontal distribution (spanned locally by the coordinate tetrads $v_\alpha$).

The nonphysical cosmological constant, which is the scalar curvature (denoted as $R_G$) of the Lie group G occurring in the Lagrangian of the usual Kaluza-Klein model [5], [7], [9], is absent in the Lagrangian (3.22) because in formula (3.24) we restricted $R_v$ to the four dimensional horizontal distribution spanned by the $v_K$.

For the same reason, even though the metric given in formula (3.4) is bi-invariant (i.e., both right and left invariant), the theorem does not require that the right-invariant metric to also be left-invariant, which in the usual Kaluza-Klein model restricts the choice of Lie groups [2], [10].

The complex scalar fields $\rho$ and $\mu$ generalize to maps from the Kaluza-Klein manifold M to a complex vector space V, where V is a unitary representation of the gauge group G, acting on the right side of V. Note that $\rho$ is a special solution of the Euler-Lagrange equation for the Kaluza-Klein Lagrangian (3.22) which is covariant; i.e., $\rho$ commutes with the right action of G on M and V, as required for a Yang-Mills Lagrangian [14]. Only horizontal vector fields act on the scalar fields $\rho$ and $\mu$ in the Lagrangian (3.22).

The Kaluza-Klein tetrad model has been generalized to include fermions, gravitons, and gauge bosons as components of a Kaluza-Klein metric together with covariant scalar fields [20], [22]. With the additional terms in the Lagrangian that are required for the Higgs mechanism, $\mu$ acts as a Higgs field in this model [22].

## References

1. R. Utiyama, "Invariant Theoretical Interpretation of Interaction", Physical Review 101, 1597–1607 (1956).

2. R. Kerner, "Generalization of the Kaluza-Klein theory for an arbitrary non-abelian gauge group", Ann. Inst. Henri Poincare 9, 143–152 (1968).

3. A. Trautman, "Fiber bundles associated with space-time", Rep. Math. Phys. 1, 29–62 (1970).

4. Y. M. Cho, "Higher dimensional unifications of gravitation and gauge theories", J. Math. Phys. 16, 2029–2035 (1975).

5. C. A. Orzalesi and M. Pauri, "Spontaneous compactification, gauge symmetry, and the vanishing of the cosmological constant", Phys. Lett. 107B, 186–190 (1981).

6. A. Salem and J. Strathdee, "On Kaluza-Klein theory", Ann. Phys. 141, 316–352 (1982).

7. M. W. Kalinowski, "Vanishing of the cosmological constant in non-abelian Kaluza-Klein theories", Int. J. Theor. Phys. 22, 385–396 (1983).

8. R. Percacci and S. Randjbar-Daemi, "Kaluza-Klein theories on bundles with homogeneous fibers", J. Math. Phys. 24, 807–814 (1983).

9. D. J. Toms, "Kaluza-Klein theories", <u>Workshop on Kaluza-Klein Theories</u>, Chalk River, Ontario, edited by H. C. Lee (World Scientific, Singapore, 1984), pp. 185–232.

10. C. C. Chiang, S. C. Lee, G. Marmo, and S. L. Lou, "Curvature tensor for Kaluza-Klein theories with homogeneous fibers", Phys. Rev. D 32, 1364–1368 (1985).

11. M. Gockeler and T. Schucker, <u>Differential Geometry, Gauge Theories, and Gravity</u> (Cambridge University Press, Cambridge, 1987), pp. 61–71 and 197–199.

12. R. Coquereaux and A. Jadcyzk, <u>Riemannian Geometry, Fiber Bundles, Kaluza-Klein Theories, and All That</u> (World Scientific, Singapore, 1988).

13. A. Macias, G. J. Fuentes y Martinez, and O. Obregon, "The Dirac Equation in 5-Dimensional Wesson Gravity", Gen. Rel. Grav. 25, 549–560 (1993).

14. G. L. Naber, Topology, Geometry, and Gauge Fields: Foundations (Springer-Verlag, New York, 1997).

15. L. O'Raifeartaigh, The Dawning of Gauge Theory, (Princeton University Press, Princeton, 1997), pp. 112–116 and 121–144.

16. P. S. Wesson, Space-Time-Matter, Modern Kaluza-Klein Theory (World Scientific, Singapore, 1999).

17. G. W. Ma and Z. K. Guo, "5D Dirac Equation in Induced Matter Theory", Int. J. Theor. Phys. 41, 1733–1743 (2002).

18. P. S. Wesson, Five-Dimensional Physics: Classical and Quantum Consequences of Kaluza-Klein Cosmology (World Scientific, Singapore, 2006), pp. 116–119.

19. F. Reifler and R. Morris, "Unification of the Dirac and Einstein Lagrangians in a tetrad model", J. Math. Phys. 36, 1741–1752 (1995).

20. F. Reifler and R. Morris, "Inclusion of gauge bosons in the tensor formulation of the Dirac theory", J. Math. Phys. 37, 3630–3640 (1996).

21. F. Reifler and R. Morris, "Measuring a Kaluza-Klein radius smaller than the Planck length", Phys. Rev. D 67 064006 (2003).

22. F. Reifler and R. Morris, "Geometric Origin of Physical Constants in a Kaluza-Klein Tetrad Model", Foundations of Physics Letters 19, 657– 673 (2006).

23. P. A. M. Dirac, "The electron wave equation in Riemann space", Max Planck Festschrift (1958), edited by W. Frank (Veb Deutscher Verlag der Wissenschaften, Berlin, 1958), pp. 339–344.

24. S. Weinberg, Gravitation and Cosmology (John Wiley and Sons, New York, 1972), pp. 365–373.

25. D. J. Hurley and M. A. Vandyck, Geometry, Spinors, and Applications, (Springer-Verlag, New York, 2000).

26. A. Zecca, "Dirac Equation in Space-Time with Torsion", Int. J. Theor. Phys. 41, 421–428 (2002).

27. J. W. Maluf, "Dirac spinor fields in the teleparallel gravity: Comment on "Metric-affine approach to teleparallel gravity"", Phys. Rev. D 67, 108501 (2003).

28. R. A. Mosna and J. G. Pereira, "Some Remarks on the Coupling Prescription of Teleparallel Gravity", Gen. Rel. Grav. 36, 2525–2538 (2004).

29. F. Reifler and R. Morris, "Hestenes' Tetrad and Spin Connections", Int. J. of Theor. Phys. 44, 1307–1324 (2005).

30. M. Leclerc, "Hermitian Dirac Hamiltonian in time dependent gravitational field", Class. Quant. Grav. 23, 4013–4019 (2006).

31. M. Arminjon and Frank Reifler, "Dirac equation: Representation independence and tensor transformation, arXiv:0707.1829 v1 [quant-ph] 12 July 2007.

32. D. Hestenes, "Real spinor fields", J. Math. Phys. 8, 798–808 (1967).

33. Y. Takahashi, "The Fierz identities - a passage between spinors and tensors", J. Math. Phys. 24, 1783–1790 (1983).

34. Y. Takahashi, "The Fierz identities", Progress in Quantum Field Theory, edited by E. Ezawa and S. Kamefuchi (Elsevier Science Publishers, Amsterdam, 1986), pp. 121–132.

35. Y. Takahashi, An Introduction to Field Quantization (Pergamon Press, Oxford, 1969).

36. F. Reifler and R. Morris, "Fermi quantization of tensor systems", Int. J. Mod. Phys. A 9, 5507–5515 (1994)

37. D. Hestenes, "Vectors, spinors, and complex numbers in classical and quantum physics", Amer. J. Phys. 39, 1013–1027 (1971).

38. R. Geroch, "Spinor Structure of Space-Times in General Relativity I", J. Math. Phys. 9, 1739–1744 (1968).

39. R. Geroch, "Spinor Structure of Space-Times in General Relativity II", J. Math. Phys. 11, 343–348 (1970).

40. C. J. Isham, "Spinor fields in four dimensional space-time", Proc. Royal Soc. of London, A. 364, 591–599 (1978).

41. S. J. Avis and C. J. Isham, "Lorentz gauge invariant vacuum functionals for quantized spinor fields in non-simply connected space-times", Nuclear Phys. B 156, 441–455 (1979).

42. R. Penrose and W. Rindler, Spinors and Space-time, Volume 1, (Cambridge University Press, Cambridge, 1986), pp. 55–56.

43. H. B. Lawson, Jr. and M. L. Michelsohn, Spin Geometry (Princeton University Press, Princeton, 1989), pp. 78−93.

44. G. Sardanashvily, "Covariant Spin Structure", J. Math. Phys. 39, 4874–4890 (1998).

45. Y. Aharonov and L. Susskind, "Observability of the Sign Change of Spinors under $2\pi$ Rotations", Phys. Rev. 158, 1237–1238 (1967).

46. J. Byrne, Young's double beam interference experiment with spinor and vector waves", Nature 275, 188–191 (1978).

47. M. P. Silverman, "The curious problem of spinor rotation", Eur. J. Phys. 1, 116–123 (1980).

48. G. C. Hegerfeldt and S. N. M. Ruijsenaars, "Remarks on causality, localization, and spreading of wave packets", Phys. Rev. D 22, 377–384 (1980).

49. B. Thaller and S. Thaller, "Remarks on the localization of Dirac particles", Il Nuovo Cimento 82A, 222–228 (1984).

50. F. Reifler and R. Morris, "Unobservability of bispinor two-valuedness in Minkowski space-time", Ann. Phys. 215, 264–276 (1992).

51. F. Reifler and A. Vogt, "Unique continuation of some dispersive waves", Commun. In Partial Differential Eqns. 19, 1203–1215 (1994).

52. V. A. Zhelnorovich, "Cosmological solutions of the Einstein-Dirac equations", Gravitation and Cosmology 2, 109–116 (1996).

53. V. A. Zhelnorovich, "On Dirac equations in the formalism of spin coefficients", Gravitation and Cosmology 3, 97–99 (1997).

54. J. D. Hamilton, "The Dirac equation and Hestenes' geometric algebra", J. Math. Phys. 25, 1823–1832 (1984).

55. F. Reifler and R. Morris, "Flavor symmetry of the tensor Dirac theory", J. Math. Phys. 40, 2680–2697 (1999).

56. J. D. Bjorken and S. D. Drell, Relativistic Quantum Mechanics (McGraw Hill, New York, 1964), pp. 16–26 and 66–70.

57. F. Mandl and G. Shaw, Quantum Field Theory, (John Wiley and Sons, New York, 1986), pp. 80 and 279–289.

58. P. A. M. Dirac, "The Electron Wave Equation in De-Sitter Space", Ann. Math. 36, 657–669 (1935).

59. J. Kocinski, "A five-dimensional form of the Dirac equation", J. Phys. A: Math. Gen. 32, 4257–4277 (1999).

60. M. Moreno, "Closed formula for the product of n Dirac matrices", J. Math. Phys. 26, 576–584 (1985).

61. N. N. Bogoliubov and D. V. Shirkov, Quantum Fields (Benjamin/Cummings, London, 1983), pp. 46 – 47.

62. D. E. Soper, Classical Field Theory (John Wiley and Sons, New York, 1976), pp. 116-117 and 222-225.

63. M. P. DoCarmo, Riemannian Geometry (Birkhauser, Boston, 1992), pp. 40–41.